\documentstyle[floats,aps,prd,epsfig,amsmath]{revtex}

\newcommand{\sgn}[1]{{\,\operatorname{sgn}\left(#1\right)}}
\newcommand{\sqz}{\epsilon}

\begin{document}

\title{Closed cosmologies with a perfect fluid and a scalar field}
\author{Alan Coley\footnote{Department of Mathematics and Statistics,
    Dalhousie University, Halifax, Nova Scotia, B3H 3J5, Canada. E-mail:
    aac@mscs.dal.ca} \ and 
  Martin Goliath\footnote{Department of Physics, Stockholm University,
    Box 6730, S-113 85 Stockholm, Sweden. E-mail: goliath@physto.se}} 
\maketitle

\begin{abstract}
Closed, spatially homogeneous cosmological models with a perfect
fluid and a scalar field with exponential potential are investigated, 
using dynamical systems methods. First, we consider the closed 
Friedmann-Robertson-Walker models, discussing the global dynamics in 
detail. Next, we investigate Kantowski-Sachs models, for which the future 
and past attractors are determined. The global asymptotic behaviour of
both the Friedmann-Robertson-Walker and the Kantowski-Sachs models is
that they either expand from an initial singularity, reach a maximum
expansion and thereafter recollapse to a final singularity
(for all values of the potential parameter $\kappa$), or else they expand
forever towards a flat power-law inflationary solution
(when $\kappa^2<2$). As an illustration of the intermediate dynamical
behaviour of the Kantowski-Sachs models, we examine the cases of no
barotropic fluid, and of a massless scalar field in detail. We also
briefly discuss Bianchi type IX models. 
\end{abstract}

\section{Introduction}

Cosmological models with a scalar field and an exponential potential
are of fundamental importance in the study of the universe. These
models are motivated by the fact that they arise naturally in
alternative theories of gravity and occur as the low-energy limit in
supergravity theories \cite{T3,T5}. By using qualitative techniques,
the well-known power-law inflationary solution has been shown to be an
attractor for all initially expanding Bianchi models (except a
subclass of the Bianchi type IX models which will recollapse) in the
class of spatially homogeneous Bianchi models
\cite{R9,ColeyIbanezVanDenHoogen}. More recently, cosmological models
which contain both barotropic matter and a scalar field with an
exponential potential have been studied \cite{bchio}, partially
motivated by the fact that there exist spatially flat isotropic
scaling solutions in which the energy density due to the scalar field
is proportional to the energy density of the perfect fluid
\cite{CopelandLiddleWands}. In \cite{BillyardColeyVanDenHoogen} the
stability of these cosmological scaling solutions within the class of
spatially homogeneous cosmological models with a perfect fluid subject
to the equation of state $p=(\gamma-1)\mu$ (where $\gamma$ is a
constant satisfying $0<\gamma<2$) was studied and it was found that
when $\gamma>2/3$, and particularly for realistic matter with
$\gamma\ge1$, the scaling solutions are unstable; essentially they are
unstable to curvature perturbations, although they are stable to shear
perturbations.  In addition, in \cite{HCW} homogeneous and isotropic
spacetimes with non-zero spatial curvature were studied. 

It is clearly of interest to study more general cosmological models.
One class of models of particular interest are those with
{\it positive spatial curvature}. These models have attracted less
attention since they are more complicated
mathematically. Positive-curvature Friedmann-Robertson-Walker (FRW)
models \cite{HCW,R3,WE,Aber}, Kantowski-Sachs models \cite {R5} and
Bianchi type IX models \cite{R9,WE,Uggla-zurMuhlen,HO} have been
studied using qualitative methods, although rigorous analyses using a
new set of compact variables have not been carried out. The Bianchi
type IX models are known to have very complicated dynamics, exhibiting the
characteristics of chaos \cite{WE,chaos}, and are hence beyond the
scope of the present study. Recently \cite{G} positive-curvature FRW
models and Kantowski-Sachs models with a perfect fluid and a
cosmological constant have been investigated using qualitative methods
and utilizing compactified variables. 

The outline of this paper is as follows. We shall first
comprehensively study the qualitative properties of the class of
positive-curvature FRW models with a barotropic fluid and a
non-interacting scalar field with an exponential potential, extending
and generalising work by Turner \cite{turner}, who used different
basic variables. We shall then analyse the qualitative properties of
the Kantowski-Sachs models. Positive-curvature FRW models and
Kantowski-Sachs models belong to the class of spherically symmetric
models, and hence the present work is a natural extension of recent
work \cite{Bogoyavlensky,ccgnu}. Indeed, it turns out that
understanding the dynamics of the Kantowski-Sachs models is crucial
for understanding the global dynamics of general spherically symmetric
similarity models \cite{CG}. 

\newpage

\subsection{Matter model}

The matter content of the models is taken to be a perfect fluid and 
a scalar field with exponential potential. The corresponding 
energy-momentum tensor is
\begin{eqnarray}
  T_{ab}&=&(T_{{\rm pf}\,ab}+T_{{\rm sf}\,ab}) , \\
  T_{{\rm pf}\,ab}&=&\mu u_au_b+p\left(u_au_b+g_{ab}\right) , \\
  T_{{\rm sf}\,ab}&=&\phi_{,a}\phi_{,b}-
  \left(\frac{1}{2}\phi_{,c}\phi^{,c}+V(\phi)\right)g_{ab} , \\
  V(\phi)&=&V_0e^{-\kappa\phi} , 
\end{eqnarray}
where $\kappa$ is a non-negative constant, and the pressure is given
by $p=(\gamma-1)\mu$, with the equation-of-state parameter in the
range $1\leq\gamma\leq2$. The fluid energy density $\mu$ 
and the scalar field $\phi$ are functions of a timelike coordinate $t$. A
dot denotes differentiation with respect to $t$ and throughout
units are used in which $c = 8 \pi G = 1$. The matter components are 
assumed to be non-coupled, and thus they are separately 
conserved:
\begin{eqnarray}
  \nabla_aT_{\rm pf}^{ab}=&0&=\nabla_aT_{\rm sf}^{ab} .
\end{eqnarray}
For convenience, we define
\begin{equation}
  X=\frac{1}{\sqrt{2}}\dot{\phi} . 
\end{equation}

\section{Closed Friedmann models}

We start our investigation of closed cosmological models with a
perfect fluid and a scalar field by looking at the closed FRW models.
The line element for these models can be written
\begin{equation}
  ds^2=-dt^2+S(t)^2dr^2+S(t)^2\sin^2r\,d\Omega^2 .
\end{equation}
The expansion of the fluid congruence is given by $\theta=3\dot{S}/S$, and
the evolution equation for the curvature $K\equiv9/(\theta^2S^2)$ is
\begin{equation}
  \dot{K}=-\frac{2K}{3\theta}(3\dot{\theta}+\theta^2) .
\end{equation}
The conservation equations yield
\begin{eqnarray}
  \dot{\mu}&=&-\gamma\theta\mu , \\
  \dot{X}&=&-\theta X+\frac{\kappa}{\sqrt{2}}V .
\end{eqnarray}
From the field equations we obtain
\begin{eqnarray}
  \mu&=&\frac{1}{3}\left[(1+K)\theta^2-3X^2-3V\right]
  \label{eq:FriedmannFRW},\\ 
  \dot{\theta}&=&-\frac{1}{3}\left\{\theta^2+\frac{1}{2}D^2+
  \frac{3}{2}\left[3X^2-3V+3(\gamma-1)\mu\right]\right\} .
\end{eqnarray}
Assuming $\mu\geq0$, the Friedmann equation,
Eq. (\ref{eq:FriedmannFRW}), 
shows that $D=\sqrt{(1+K)\theta^2}$ is a dominant
quantity. Thus, compact variables can be defined according to 
\begin{equation}
  Q_0=\frac{\theta}{D} , \quad
  U=\frac{\sqrt{3}X}{D} , \quad
  W=\frac{\sqrt{3V}}{D} .
\end{equation}
Note also that the curvature is given by
\begin{equation}
  K=\frac{1-Q_0^2}{Q_0^2} .
\end{equation}
The Friedmann equation becomes
\begin{equation}
  \Omega_D=\frac{3\mu}{D^2}=1-U^2-W^2 . 
\end{equation}
Defining a new independent variable, $'=d/d\tau=\frac{3}{D}\,d/dt$, the 
evolution equation for $D$
\begin{eqnarray}
  D'&=&-3Q_0\left(U^2+\frac{\gamma}{2}\Omega_D\right)D 
\end{eqnarray}
decouples. Thus, a reduced set of evolution equations is obtained:
\begin{eqnarray}
  Q_0'&=&(1-Q_0^2)\left[1-3\left(U^2+\frac{\gamma}{2}\Omega_D\right)\right] 
  , \nonumber\\
  U'&=&3Q_0U\left[-1+\left(U^2+\frac{\gamma}{2}\Omega_D\right)\right]+
  \sqrt{\frac{3}{2}}\kappa W^2 , \label{eq:dynsysFRW}\\
  W'&=&3Q_0\left(U^2+\frac{\gamma}{2}\Omega_D\right)W -
  \sqrt{\frac{3}{2}}\kappa UW \nonumber.
\end{eqnarray}
There is also an auxiliary evolution equation
\begin{equation}\label{eq:Dprime}
  \Omega_D'=-3Q_0\left[(1-\Omega_D)\gamma-2U^2\right]\Omega_D ,
\end{equation}
and it is straight-forward to consider the set of variables 
$(Q_0,U,\Omega_D)$, rather 
than $(Q_0,U,W)$ \cite{turner}. Note that by setting $\kappa=0,U=0$, 
and identifying $\Lambda=V_0$, $\Omega_\Lambda=W^2$,
the evolution equations corresponding to closed FRW models with a cosmological 
constant are obtained \cite{G}.

It is also useful to consider the deceleration parameter, given by
\begin{equation}
  q_{\rm pf}\equiv-\left(1+3\frac{u^a_{\rm pf}\nabla_a\theta_{\rm pf}}
  {\theta_{\rm pf}^2}\right)=
  -\frac{1}{Q_0^2}
  \left[1-3\left(U^2+\frac{\gamma}{2}\Omega_D\right)\right] ,
\end{equation}
for $Q_0\ne 0$. From this expression, we can see that there is an inflationary
region ($q_{\rm pf}<0$) in the state space whenever
$\Omega_D<\frac{2}{3\gamma}(1-3U^2)$. However, as will be seen below,
it is only when $\kappa^2<2$ that there exist attractors that are
inflationary. 
Note also that for $Q_0\neq0$, 
\begin{equation}
Q_0' = -(1-Q_0^2)Q_0^2 q_{\rm pf},
\end{equation}
so that $Q_0'<0$ whenever $q_{\rm pf}>0$ in which case $Q_0$ is itself
monotonic. When $q_{\rm pf}<0$, that is in the inflationary region,  $Q_0$
need not be monotonic -- for example, see the orbits close to $_+\Phi$ in 
Figs. \ref{fig:FRW0} and \ref{fig:FRW1}.

The dynamical system Eqs. (\ref{eq:dynsysFRW}) is symmetric
under the transformation 
\begin{equation}\label{eq:symmFRW}
  \left(\tau,Q_0,U,W\right) \rightarrow \left(-\tau,-Q_0,-U,W\right) .
\end{equation}
Thus, it is sufficient to discuss the behaviour in one part of the state
space, the dynamics in the other part being obtained by
Eq. (\ref{eq:symmFRW}).

Furthermore, note that
\begin{eqnarray}
  M&=&\frac{\Omega_D}{1-Q_0^2} \nonumber \\
  M'&=&-(3\gamma-2)Q_0 M
\end{eqnarray}
is a monotonic function in the regions $Q_0<0$ and $Q_0>0$ for $\Omega_D\ne0$.
As there are no equilibrium points with $Q_0=0$ when $\gamma>2/3$, $M$
acts as a monotonic function in the interior of the state space.
Consequently there can be no periodic or recurrent orbits in the interior 
state space and global results can be deduced. In addition, from the 
expression for the monotonic function $M$ we can see immediately that 
either $Q_0^2 \rightarrow 1$ or $\Omega_D \rightarrow 0$ asymptotically.

\subsection{Equilibrium points of the closed FRW dynamical system}
A number of equilibrium points can be found for the dynamical system,
Eqs. (\ref{eq:dynsysFRW}). 
In what follows, $\sqz=\pm1$ denotes the sign of $Q_0$, 
whereas $\Omega_\phi=U^2+W^2$ is a density parameter associated with
the scalar field. In Table \ref{tab:equiFRW}, the various equilibrium points
are summarized. The subscripts on the labels have the following
significance: The left subscript gives the sign of $Q_0$ and indicates
whether the corresponding model is expanding (+) or contracting
($-$). The right subscript gives the sign of $U$; i.e., the sign of
$\dot{\phi}$. 

\begin{table}
  \begin{center}
    \begin{tabular}{ll|ccc|l}
      & Interpretation & $Q_0$ & $U$ & $W$ & Note \\ \hline

      $_\pm{\rm F}$ & Flat Friedmann & 
      $\sqz$ & 0 & 0 & \\

      $_\pm{\rm K}_\pm$ & Kinetic dom. &
      $\sqz$ & $\pm1$ & 0 & \\

      $_\pm\Phi$ & Scalar-field dom. &
      $\sqz$ & $\frac{\kappa}{\sqrt{6}}\sqz$ & $\sqrt{1-\frac{\kappa^2}{6}}$ &
      $\kappa^2<6$ \\

      $_\pm{\rm X}$ & Curvature scaling &
      $\frac{\kappa}{\sqrt{2}}\sqz$ & $\frac{\sqz}{\sqrt{3}}$ & 
      $\sqrt{\frac{2}{3}}$ &
      $\kappa^2<2$ \\

      $_\pm{\rm FS}$ & Flat matter scaling &
      $\sqz$ & $\sqrt{\frac{3}{2}}\frac{\gamma}{\kappa}\sqz$ & 
      $\sqrt{\frac{3}{2}}\frac{1}{\kappa}\sqrt{\gamma(2-\gamma)}$ &
      $\kappa^2>3\gamma$ \\

      ${\rm S}_\pm$ & Static &
      0 & $\pm\sqrt{-\frac{3\gamma-2}{3(2-\gamma)}}$ & 0 &
      $\gamma<2/3$\\ 
    \end{tabular}
    \caption{Equilibrium points of the closed FRW models.}\label{tab:equiFRW}
  \end{center}
\end{table}

The equilibrium points labeled $_\pm{\rm F}$ correspond to the flat Friedmann
solution. For these points, the scalar field vanishes ($U=0=W$).
There is an orbit from $_+{\rm F}$ to $_-{\rm F}$ and this
orbit represents the closed FRW solution with no scalar field,
starting from a Big Bang at $_+{\rm F}$ and recollapsing to a
``Big Crunch'' at $_-{\rm F}$. 

The K points represent exact solutions with a massless scalar field
($W=0$). As the fluid is negligible ($\Omega_D=0$), these solutions
are dominated by the kinetic term $U$. They correspond to Jacobs analogues
of Kasner solutions in which $U$ takes on the role of a shearing mode
\cite{WE}. 

There are equilibrium points  with non-vanishing potential, where the
scalar field dominates ($\Omega_D=0$). These points $_\pm\Phi$ are only
physical when $\kappa^2<6$. For $\kappa^2=6$, $_+\Phi$ coincides with
$_+{\rm K}_+$, and $_-\Phi$ with $_-{\rm K}_-$. For $\kappa^2>6$,
they are outside the physical part of the state space. The equilibrium
point with $\epsilon=+1$ (i.e., $_+\Phi$) is a sink, and for
$\kappa^2<2$ it corresponds to the power-law inflationary attractor
solution (cf. the expression for $q_{\rm pf}$ in Table \ref{tab:qFRW}).

There are also points $_\pm{\rm X}$ for which the matter is
unimportant, but the curvature is non-vanishing ($Q_0^2\neq1$), and
tracks the scalar field. The corresponding solutions are called
curvature scaling solutions \cite{HCW}. These solutions only exist
when $\kappa^2<2$. For $\kappa^2=2$, $_+{\rm X}$ coincides with
$_+\Phi$, and $_-{\rm X}$ with $_-\Phi$. Above this value of $\kappa$,
these equilibrium points are outside the physical part of the state
space. 

When $\kappa^2>3\gamma$, there is a flat matter scaling solution, for which
both the fluid and the scalar field are dynamically important. The
corresponding equilibrium points are denoted $_\pm{\rm FS}$, and for
$\kappa^2=3\gamma$ they coincide with $_\pm\Phi$.

Finally, there are equilibrium points ${\rm S}_\pm$ corresponding to static
solutions, analogous to the Einstein static universe. These are only
physical when $\gamma<2/3$. 
For $\gamma=2/3$, a set of equilibrium points appears along the line
$U=0=W$, signaling a change of stability when the points ${\rm S}_\pm$
leave the physical state space. In what follows, we will only consider
equations of state for which $1\leq\gamma\leq2$, hence we will not
consider these points further.

\begin{table}
  \begin{center}
    \begin{tabular}{l|ccc}
      & $\Omega_D$ & $\Omega_\phi$ & $q_{\rm pf}$ \\ \hline
      
      $_\pm{\rm F}$ & 1 & 0 & $\frac{1}{2}(3\gamma-2)$ \\

      $_\pm{\rm K}_\pm$ & 0 & 1 & 2 \\

      $_\pm\Phi$ & 0 & 1 & $\frac{1}{2}(\kappa^2-2)$ \\

      $_\pm{\rm X}$ & 0 & 1 & 0 \\

      $\pm{\rm FS}$ & $\frac{1}{\kappa^2}(\kappa^2-3\gamma)$ & 
      $\frac{3\gamma}{\kappa^2}$ & $\frac{1}{2}(3\gamma-2)$ \\

      ${\rm S}_\pm$ & $\frac{4}{3(2-\gamma)}$ &
      $-\frac{3\gamma-2}{3(2-\gamma)}$ & Not defined \\ 
    \end{tabular}
    \caption{The physical quantities $\Omega_D$, $\Omega_\phi$, and
      $q_{\rm pf}$ for the different equilibrium points of the closed
      FRW models.}\label{tab:qFRW} 
  \end{center}
\end{table}

\begin{table}
  \begin{center}
    \begin{tabular}{l|ccc}
      & \multicolumn{3}{c}{Eigenvalues} \\ \hline

      $_\pm{\rm F}$ & 
      $(3\gamma-2)\sqz$ &
      $-\frac{3}{2}(2-\gamma)\sqz$ &
      $\frac{3\gamma}{2}\sqz$ \\

      $_\pm{\rm K}_\pm$ &
      $4\sqz$ &
      $3(2-\gamma)\sqz$ &
      $3\sqz-\sqrt{\frac{3}{2}}\kappa\sgn{U}$ \\

      $_\pm\Phi$ &
      $-(3\gamma-\kappa^2)\sqz$ &
      $-\frac{1}{2}(6-\kappa^2)\sqz$ &
      $-(2-\kappa^2)\sqz$ \\

      $_\pm{\rm X}$ &
      $-\frac{\kappa}{\sqrt{2}}(3\gamma-2)\sqz$ &
      \multicolumn{2}{c}
      {$\frac{1}{\sqrt{2}}(-\kappa\sqz\pm\sqrt{8-3\kappa^2})$} \\ 

      $\pm{\rm FS}$ &
      $(3\gamma-2)\sqz$ &
      \multicolumn{2}{c}{
        $-\frac{3}{4}\left[(2-\gamma)\pm\frac{1}{\kappa}
        \sqrt{(2-\gamma)[24\gamma^2-(9\gamma-2)\kappa^2]}\right]\sqz$ } \\

      ${\rm S}_\pm$ &
      $-\frac{\kappa}{\sqrt{2}}
      \sqrt{-\frac{3\gamma-2}{2-\gamma}}\sgn{U}$ &
      \multicolumn{2}{c}
      {$\pm\sqrt{2}\sqrt{-(3\gamma-2)}$} \\ 
    \end{tabular}
    \caption{Eigenvalues for the different equilibrium points of the
      closed FRW models.}\label{tab:eigenFRW}
  \end{center}
\end{table}

\begin{table}
  \begin{center}
    \begin{tabular}{lll}
      \multicolumn{3}{c}{Past attractors} \\ \hline
      Expanding from a singularity ($Q_0>0$)
      & $_+{\rm K}_-$ & Always \\
      & $_+{\rm K}_+$ & $\kappa^2<6$ \\ \hline
      Contracting from a dispersed state ($Q_0<0$)
      & $_-\Phi$ & $\kappa^2<2$ (and $\kappa^2<3\gamma$) \\ \hline\hline
      \multicolumn{3}{c}{Future attractors} \\ \hline
      Contracting to a singularity ($Q_0<0$)
      & $_-{\rm K}_+$ & Always \\
      & $_-{\rm K}_-$ & $\kappa^2<6$ \\ \hline
      Expanding to a dispersed state ($Q_0>0$)
      & $_+\Phi$ & $\kappa^2<2$ (and $\kappa^2<3\gamma$) \\ 
    \end{tabular}
    \caption{Summary of sources and sinks for the closed FRW
      models.}\label{tab:stabFRW} 
  \end{center}
\end{table}

\begin{figure}
  \centerline{\hbox{\epsfig{figure=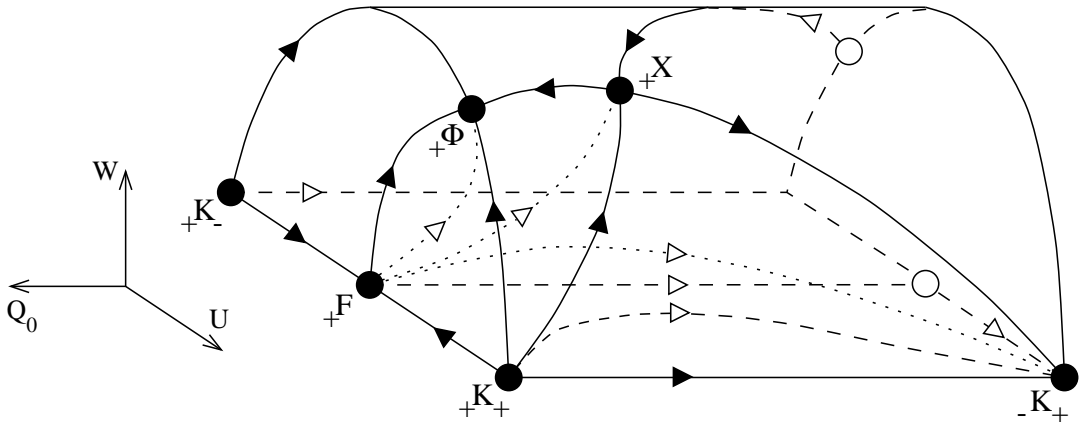,width=0.8\textwidth}}}
  \caption{The state space for closed FRW models with a scalar
    field when $\kappa^2<2/3$. Dashed curves and white arrows and
  circles are screened. Dotted orbits are in the interior of the state
  space.}\label{fig:FRW0} 
\end{figure}

Table \ref{tab:qFRW} presents some physical quantities for the various
equilibrium points, and Table \ref{tab:eigenFRW} lists their eigenvalues.
The sources and sinks of the dynamical system when $\gamma>2/3$ are
listed in Table 
\ref{tab:stabFRW}, and the global behaviour for different values of
$\kappa$ can be summarized as follows:
The state space when $0<\kappa^2<2/3$ is depicted in Fig. \ref{fig:FRW0},
where the features in the rear part of the state space have been
suppressed. Dashed and full curves represent orbits in the boundary
submanifolds, while dotted curves represent orbits in the interior.
Both $_+{\rm K}_-$ and $_+{\rm K}_+$ are past attractors, while
$_-{\rm K}_+$ and $_+\Phi$ act as future attractors. There are also
orbits from $_+{\rm K}_-$, whose future attractor is $_-{\rm K}_-$ at
the rear of the figure. Note that orbits future asymptotic to $_+\Phi$
correspond to solutions that exhibit power-law inflation
($-1<q_{\rm pf}<0$, see Table \ref{tab:qFRW}).
Observe that the outgoing eigenvector directions from the saddle point
$_+{\rm F}$ span a separatrix surface in the interior of the state
space. Similarly, $_+{\rm X}$ is a saddle for which the ingoing
eigenvector directions span another separatrix surface. These
separatices confine orbits in the interior state space to specific
regions. For example, there is one region where all orbits are past
asymptotic to $_+{\rm K}_+$ and future asymptotic to $_-{\rm K}_+$.

\begin{figure}
  \centerline{\hbox{\epsfig{figure=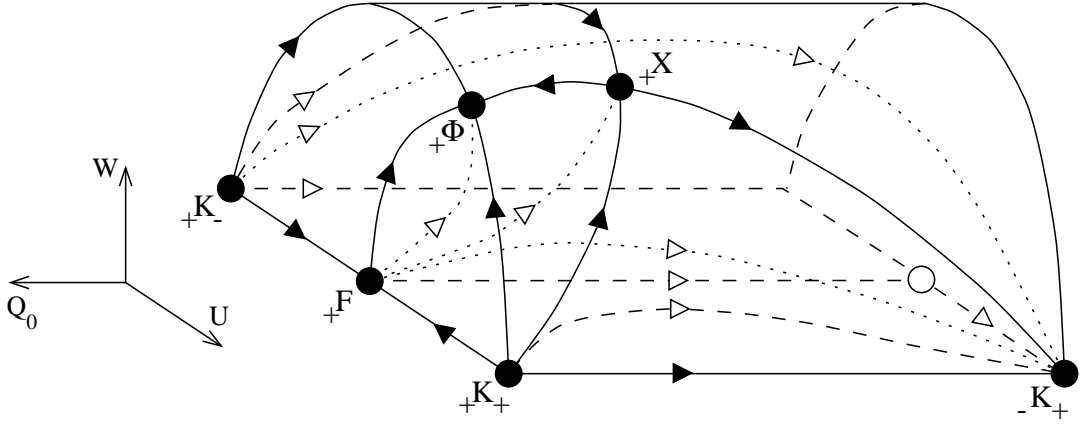,width=0.8\textwidth}}}
  \caption{The state space for closed FRW models with a scalar
    field when $2/3<\kappa^2<2$. Dashed curves and white arrows and
  circles are screened. Dotted orbits are in the interior of
    the state space.}\label{fig:FRW1} 
\end{figure}

\begin{figure}
  \centerline{\hbox{\epsfig{figure=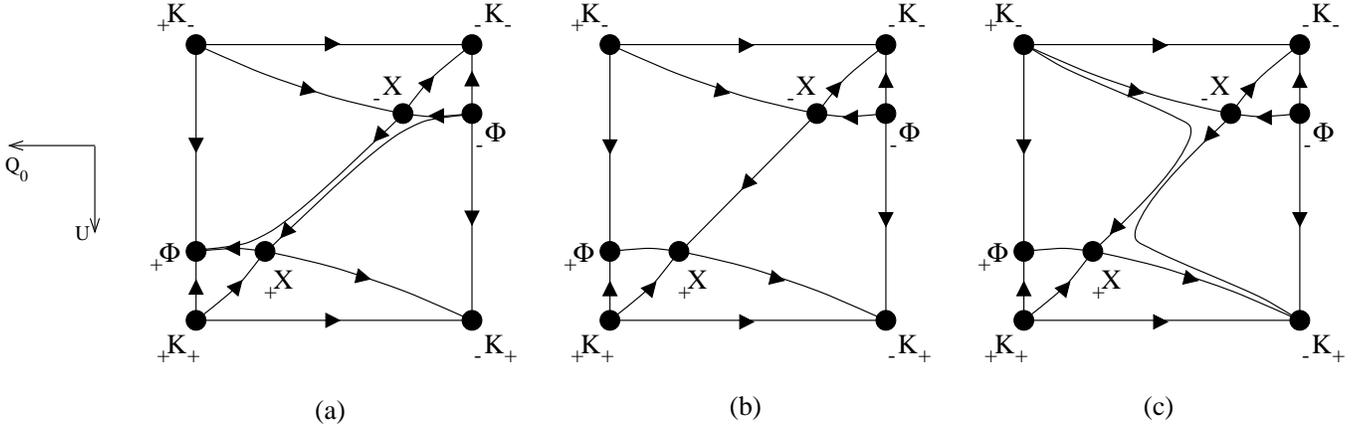,width=1.0\textwidth}}}
  \caption{Separatrix orbits in the $\Omega_D=0$ submanifold of the
    closed FRW models when
    $\kappa^2$ is near $2/3$. (a)
    $\kappa^2<2/3$; (b) $\kappa^2=2/3$; (c)
    $\kappa^2>2/3$.}\label{fig:vacuumsep}  
\end{figure}

When $\kappa^2>2/3$, the separatrix surface associated with
$_+{\rm X}$ changes structure, see Fig. \ref{fig:FRW1}. This is easiest seen
by considering the separatrix orbits in the $\Omega_D=0$ submanifold, see
Fig. \ref{fig:vacuumsep}. When $\kappa^2<2/3$, there is one separatrix
orbit from $_-\Phi$ to $_+{\rm X}$, and one from $_-{\rm X}$ to
$_+\Phi$. For $\kappa^2=2/3$, these two orbits coalesce into a single
orbit from $_-{\rm X}$ to $_+{\rm X}$. This orbit corresponds to a
special ``Bouncing Universe'' solution, existing only for this
particular value of $\kappa$. When $\kappa^2>2/3$, the separatrix
orbit to $_+{\rm X}$ starts at $_+{\rm K}_-$, and the orbit from
$_-{\rm X}$ goes to $_-{\rm K}_+$. Thus, there is a bifurcation for
$\kappa^2=2/3$; however, we note that there is no stability change of
equilibrium points involved. A similar behaviour of separatrix
surfaces has been found for Bianchi type IX models \cite{Uggla-zurMuhlen}.

\begin{figure}
  \centerline{\hbox{\epsfig{figure=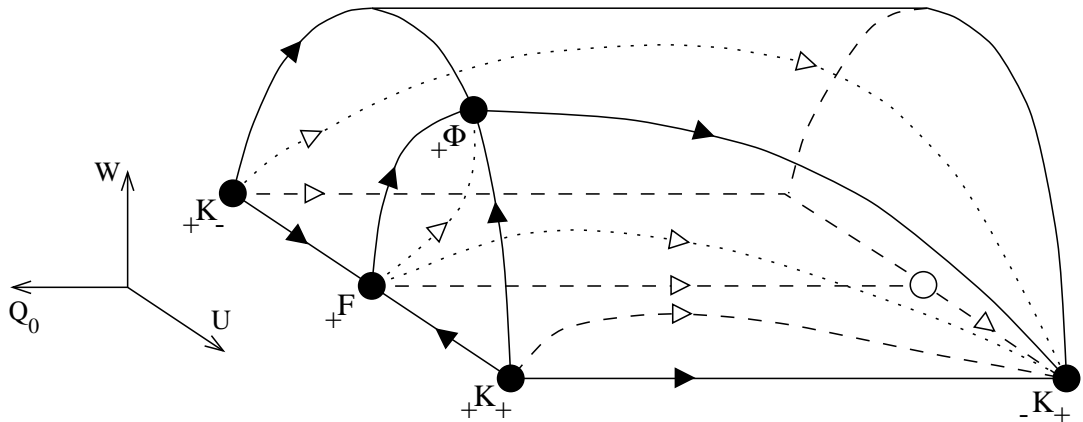,width=0.8\textwidth}}}
  \caption{The state space for closed FRW models with a scalar
    field when $2<\kappa^2<3\gamma$. Dashed curves and white arrows and
  circles are screened. Dotted orbits are in the interior
    of the state space.}\label{fig:FRW2}  
\end{figure}

When $\kappa$ increases, the equilibrium point $_+{\rm X}$ approaches
$_+\Phi$. For $\kappa^2=2$ these two points coincide, and the state
space for $\kappa^2>2$ is depicted in Fig. \ref{fig:FRW2}. 
Both $_+{\rm K}_-$ and $_+{\rm K}_+$ still act as past attractors,
while $_-{\rm K}_+$ is a future attractor. However, the stability of
$_+\Phi$ has changed; it has become a saddle. There is still a
separatrix surface associated with $_+{\rm F}$. Thus, orbits having
$_+{\rm K}_+$ as their past attractor all end at $_-{\rm K}_+$.

\begin{figure}
  \centerline{\hbox{\epsfig{figure=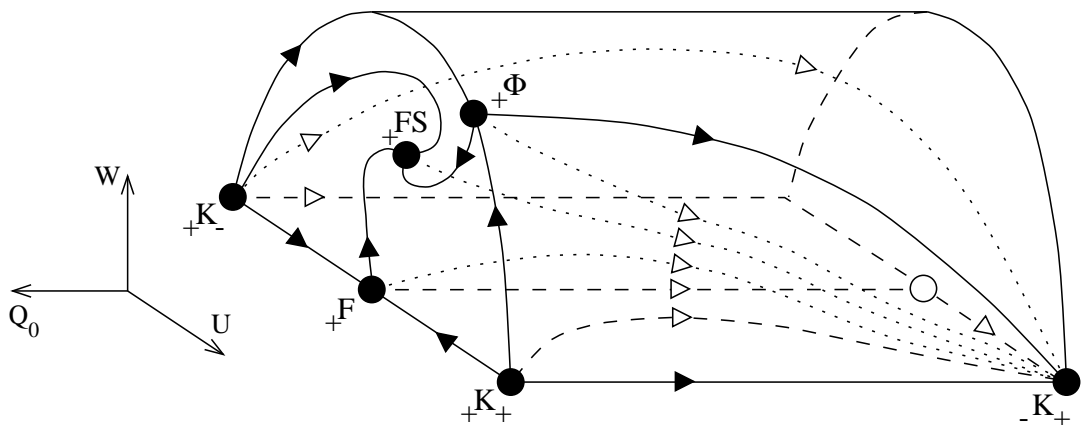,width=0.8\textwidth}}}
  \caption{The state space for closed FRW models with a scalar
    field when $3\gamma<\kappa^2<6$. Dashed curves and white arrows and
  circles are screened. Dotted orbits are in the interior of the state
  space.}\label{fig:FRW3} 
\end{figure}

For $\kappa^2>3\gamma$, the equilibrium point $_+{\rm FS}$,
corresponding to the matter-scaling solution, appears from
$_+\Phi$, see Fig. \ref{fig:FRW3}. The point $_+{\rm FS}$ is a spiral
sink with an out-going eigenvector direction entering the interior
state space. Note that this equilibrium point thus is stable in the
flat ($Q_0=1$) submanifold, but unstable to curvature perturbations
(i.e., perturbations in the $Q_0$ direction). The scalar-field
dominated point $_+\Phi$ still is a saddle, and now there is a
separatrix surface spanned by the out-going eigenvector directions there. Thus,
there are two separatrix surfaces, both of which are spiraling around the
out-going eigenvector direction of $_+{\rm FS}$.

\begin{figure}
  \centerline{\hbox{\epsfig{figure=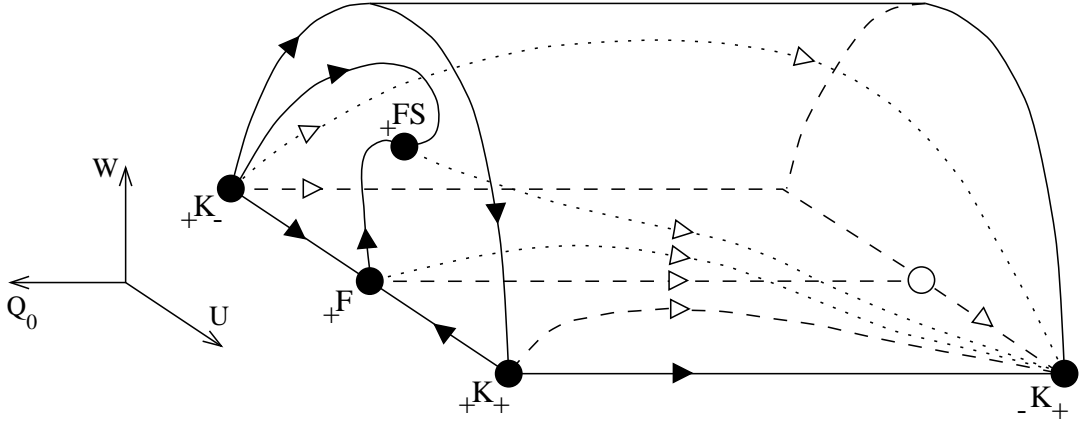,width=0.8\textwidth}}}
  \caption{The state space for closed FRW models with a scalar
    field when $\kappa^2>6$. Dashed curves and white arrows and
  circles are screened. Dotted orbits are in the interior of the state
  space.}\label{fig:FRW4} 
\end{figure}

When $\kappa$ increases, the $_+\Phi$ equilibrium point comes closer
and closer to $_+{\rm K}_+$, and for $\kappa^2=6$ they coincide. The
state space when $\kappa^2>6$ is given in Fig. \ref{fig:FRW4}. There is
only one past and one future attractor, namely $_+{\rm K}_-$ and
$_-{\rm K}_+$, respectively.

To {\bf summarize}, when $\kappa^2>2$ all solutions start from and
recollape to a singularity ($K\rightarrow K$). Thus, in this case 
solutions can neither expand forever nor inflate. When $\kappa^2<2$, 
there are also ever-expanding (${\rm K}\rightarrow\Phi$)
(and ever-collapsing $\Phi\rightarrow{\rm K}$) solutions in addition
to the recollapsing solutions. 
Inflation occurs when $\ddot{S}>0$, i.e., $3\dot{\theta}+\theta^2>0$,
which leads to the condition
\begin{equation}
  -(3\gamma-2) -3(2-\gamma)U^2 +3\gamma W^2 >0 .
\end{equation}
This corresponds to a parabolic region along the ridge of the state spaces,
Figs. \ref{fig:FRW0}, \ref{fig:FRW1} and \ref{fig:FRW2} --
\ref{fig:FRW4}. The only equilibrium points within this region are
$_\pm\Phi$ for $\kappa^2<2$, corresponding to power-law
inflation. Consequently, in the case $\kappa^2<2$ there is a subclass
of solutions that inflate.

\section{Kantowski-Sachs models}

We now turn our attention to the Kantowski-Sachs (KS) models.
The line element can be written
\begin{equation}
  ds^2=-dt^2+D_1(t)^2dx^2+D_2(t)^2d\Omega^2 ,
\end{equation}
where
\begin{equation}
  D_1=\exp\left[\beta^0(t)-2\beta^+(t)\right] , \quad 
  D_2 = \exp\left[\beta^0(t)+\beta^+(t) \right] . 
\end{equation}
The kinematic quantities of the fluid congruence are related to the Misner 
variables ($\beta^0$, $\beta^+$) by
\begin{equation}
  \theta=3\dot{\beta}^0 , \quad 
  \sigma_+=3\dot{\beta}^+ ,
\end{equation}
and the evolution equations for the metric functions $B_1\equiv D_1^{-1}$ 
and $B_2\equiv D_2^{-1}$ become
\begin{equation}
  \dot{B}_1=-\frac{1}{3}(\theta+\sigma_+)B_1 , \quad
  \dot{B}_2=-\frac{1}{3}(\theta+\sigma_+)B_2 .
\end{equation}
The conservation equations give
\begin{eqnarray}
  \dot{\mu}&=&-\gamma\theta\mu , \\
  \dot{X}&=&-\theta X+\frac{\kappa}{\sqrt{2}}V , 
\end{eqnarray}
and the field equations yield
\begin{eqnarray}
  \mu&=&\frac{1}{3}\left(\theta^2-\sigma_+^2+3B_2^2-3X^2-3V\right) , 
  \label{eq:FriedmannKS} \\
  \dot{\theta}&=&-\frac{1}{3}
  \left(\theta^2+2\sigma_+^2+6X^2-3V+\frac{3}{2}(3\gamma-2)\mu
  \right) ,\\ 
  \dot{\sigma}_+&=&\frac{1}{3}
  \left(\theta^2-\sigma_+^2-3\theta\sigma_+-3X^2-3V-3\mu\right) .
\end{eqnarray}
The Friedmann equation, Eq. (\ref{eq:FriedmannKS}), together with the 
assumption $\mu\geq0$ shows that $D=\sqrt{\theta^2+3B_2^2}$
is a dominant quantity. Consequently, compact variables are introduced 
according to
\begin{equation}
  Q_0=\frac{\theta}{D}, \quad 
  Q_+=\frac{\sigma_+}{D} , \quad
  U=\frac{\sqrt{3}X}{D}, \quad 
  W=\frac{\sqrt{3V}}{D} .
\end{equation}
The curvature variable $K=3B_2^2\,\theta^{-2}=(1-Q_0^2)/Q_0^2$
shows that the flat solutions correspond to $Q_0^2=1$. The Friedmann
equation becomes 
\begin{equation}
  \Omega_D=\frac{3\mu}{D^2}=1-Q_+^2-U^2-W^2 . 
\end{equation}
By introducing a new independent variable, $\tau$, where
$'=d/d\tau=\frac{3}{D}\,d/dt$, the evolution equation for $D$,
\begin{equation}
  D'=-\left[Q_+(1-Q_0^2)+
  3Q_0\left(Q_+^2+U^2+\frac{\gamma}{2}\Omega_D\right)\right]D , 
\end{equation}
decouples, and a reduced set of evolution equations is obtained:
\begin{eqnarray}
  Q_0'&=&(1-Q_0^2)\left[1+Q_0Q_+-
  3\left(Q_+^2+U^2+\frac{\gamma}{2}\Omega_D\right)\right] , \nonumber \\
  Q_+'&=&-(1-Q_0^2)(1-Q_+^2)+
  3Q_0Q_+\left[-1+\left(Q_+^2+U^2+\frac{\gamma}{2}\Omega_D\right)\right] , 
  \label{eq:dynsysKS} \\
  U'&=&U\left\{(1-Q_0^2)Q_+ + 3Q_0
  \left[-1+\left(Q_+^2+U^2+\frac{\gamma}{2}\Omega_D\right)\right]\right\}+
  \sqrt{\frac{3}{2}}\kappa W^2 , \nonumber \\
  W'&=&W\left\{(1-Q_0^2)Q_+ +
  3Q_0\left(Q_+^2+U^2+\frac{\gamma}{2}\Omega_D\right)\right\}-
  \sqrt{\frac{3}{2}}\kappa UW \nonumber .
\end{eqnarray}
There is also an auxiliary evolution equation:
\begin{eqnarray}
  \Omega_D'&=&
  -\Omega_D\left\{3\gamma Q_0-2\left[Q_+(1-Q_0^2)+
  3Q_0\left(Q_+^2+U^2+\frac{\gamma}{2}\Omega_D\right)\right]\right\} 
  \label{eq:Om}.
\end{eqnarray}
Note that by setting $\kappa=0,U=0$, 
and identifying $\Lambda=V_0$, $\Omega_\Lambda=W^2$,
the evolution equation corresponding to Kantowski-Sachs models with a 
cosmological constant are obtained \cite{G}.
The deceleration parameter is given by
\begin{eqnarray}
  q_{\rm pf}&=&-\frac{1}{Q_0^2}
  \left[1-3\left(Q_+^2+U^2+\frac{\gamma}{2}\Omega_D\right)\right] .
\end{eqnarray}

Note that the dynamical system Eqs. (\ref{eq:dynsysKS}) is symmetric
under the transformation 
\begin{equation}\label{eq:symmKS}
  \left(\tau,Q_0,Q_+,U,W\right) \rightarrow 
  \left(-\tau,-Q_0,-Q_+,-U,W\right) .
\end{equation}
Thus, it is sufficient to discuss the behaviour in one part of the state
space, the dynamics in the other part being obtained by
Eq. (\ref{eq:symmKS}).

The function
\begin{eqnarray}
  M&=&Q_+^{-2(3\gamma-2)}(1-Q_0^2)^{-3(2-\gamma)}\Omega_D^4 , \\
  M'&=&2Q_+^{-1}\left[(3\gamma-2)(1-Q_0^2) +3(2-\gamma)Q_+^2\right]M
\end{eqnarray}
is monotonic in the regions $Q_+>0$ and $Q_+<0$, since $2/3<\gamma<2$. 
Noting that
\begin{equation}
 \left.Q_+'\right|_{Q_+=0}=-(1-Q_0^2) < 0,
\end{equation}
we conclude that the submanifold $Q_+=0$ is not invariant, but acts as a 
membrane. Thus, the existence of $M$ rules out any periodic or recurrent 
orbits in the 
interior of the state space and again global results are possible. From the 
expression for the monotonic
function $M$ we can immediately see that asymptotically $Q_+ \rightarrow 0$,
$Q_0^2 \rightarrow 1$ or $\Omega_D \rightarrow 0$.

\subsection{Equilibrium points of the KS dynamical system}

The dynamical system, Eqs. (\ref{eq:dynsysKS}), has several
equilibrium points, which are displayed in Table. \ref{tab:equiKS}. As before, 
$\sqz=\pm1$ denotes the sign of $Q_0$, while $\Omega_\phi\equiv U^2+W^2$.
Again, the left subscript gives the sign of $Q_0$ and indicates
whether the corresponding model is expanding or contracting. The values of
$\Omega_D$, $\Omega_\phi$, and $q_{\rm pf}$ for each of the equilibrium 
points are given in Table \ref{tab:qKS} while the eigenvalues
are displayed in Table \ref{tab:eigenKS}. Note that all of the equilibrium 
points
correspond to exact self-similar cosmological models \cite{bchio,WE}.

\begin{table}
  \begin{center}
    \begin{tabular}{ll|cccc|l}
      & Interpretation & $Q_0$ & $Q_+$ & $U$ & $W$ & Note \\ \hline

      $_\pm{\rm F}$ & Flat Friedmann &
      $\sqz$ & 0 & 0 & 0 & \\

      $_\pm{\rm K}$ & Kinetic dom. &
      $\sqz$ & $\pm\sqrt{1-U_0^2}$ & $U_0$ & 0 & 
      Ring \\

      $_\pm\Phi$ & Scalar-field dom. &
      $\sqz$ & 0 & $\frac{\kappa}{\sqrt{6}}\sqz$ & 
      $\frac{1}{\sqrt{6}}\sqrt{6-\kappa^2}$ &
      $\kappa^2<6$\\
      & (when $\kappa^2<2$) & & & & & \\

      $_\pm\Xi$ & Curvature scaling &
      $2\frac{1+\kappa^2}{4+\kappa^2}\sqz$ &
      $-\frac{2-\kappa^2}{4+\kappa^2}\sqz$ &
      $\frac{\sqrt{6}\kappa}{4+\kappa^2}\sqz$ & 
      $\frac{\sqrt{6}}{4+\kappa^2}\sqrt{2+\kappa^2}$ & 
      $\kappa^2<2$ \\

      $_\pm{\rm FS}$ & Flat matter scaling &
      $\sqz$ & 0 & $\sqrt{\frac{3}{2}}\frac{\gamma}{\kappa}\sqz$ &
      $\sqrt{\frac{3}{2}}\frac{1}{\kappa}\sqrt{\gamma(2-\gamma)}$ &
      $\kappa^2>3\gamma$ \\

      $_\pm{\rm SSKS}$ & Self-similar KS &
      $\frac{2}{3\gamma-4}\sqz$ & $\frac{3\gamma-2}{3\gamma-4}\sqz$ & 0 & 0 &
      $\gamma<2/3$ \\ 
    \end{tabular}
    \caption{Equilibrium points of the KS models.}\label{tab:equiKS}
  \end{center}
\end{table}

As for the closed FRW models, $_\pm{\rm F}$ denotes the flat Friedmann
solution. Note that the closed FRW solution without a scalar
field does not appear as a submanifold of the Kantowski-Sachs models
without a scalar field. Consequently, there is no orbit connecting
$_+{\rm F}$ with $_-{\rm F}$.

There are two sets $_\pm{\rm K}$ of vacuum ($\Omega_D=0$) equilibrium
points, parameterized by the constant $U_0$, corresponding
to kinetic dominated solutions. These sets are analogues of the
``Kasner rings'' that are present for various Bianchi models.

The flat scalar-field dominated points $_\pm\Phi$, already encountered 
for the closed FRW models, appear in the Kantowski-Sachs case as well. 
As for the closed FRW models, they are physical when $\kappa^2<6$ and 
inflationary when $\kappa^2<2$.

There are also equilibrium points $_\pm\Xi$, corresponding to
curvature scaling solutions (i.e. they have $\Omega_D=0$, $Q_0^2<1$)
which are physical when $\kappa^2<2$. They are also inflationary, but
in other respects they resemble the points $_\pm{\rm X}$ of the closed
FRW models. 

As for the closed FRW models, the equilibrium points $_\pm{\rm FS}$,
corresponding to the flat matter-scaling
solution, enter the physical part of the state space when
$\kappa^2>3\gamma$.

Finally, there are also equilibrium points corresponding to the
self-similar Kantowski-Sachs solution. This solution is only physical
when $\gamma<2/3$, and so we will not consider them further.

\begin{table}
  \begin{center}
    \begin{tabular}{l|ccc}
      & $\Omega_D$ & $\Omega_\phi$ & $q_{\rm pf}$ \\ \hline

      $_\pm{\rm F}$ & 1 & 0 & $\frac{1}{2}(3\gamma-2)$ \\

      $_\pm{\rm K}$ & 0 & $U_0^2$ & 2 \\

      $_\pm\Phi$ & 0 & 1 & $\frac{1}{2}(\kappa^2-2)$ \\ 

      $_\pm\Xi$ & 0 & $12\frac{1+\kappa^2}{(4+\kappa^2)^2}$ &
      $\frac{1}{2}\frac{\kappa^2-2}{1+\kappa^2}$ \\

      $_\pm{\rm FS}$ & $\frac{\kappa^2-3\gamma}{\kappa^2}$ & 
      $\frac{3\gamma}{\kappa^2}$ & $\frac{1}{2}(3\gamma-2)$ \\

      $_\pm{\rm SSKS}$ & $-12\frac{\gamma-1}{(3\gamma-4)^2}$ & 0 & 
      $\frac{1}{2}(3\gamma-2)$ \\ 
    \end{tabular}
    \caption{The physical quantities $\Omega_D$, $\Omega_\phi$, and
      $q_{\rm pf}$ for the different equilibrium points of the KS
      models.}\label{tab:qKS} 
  \end{center}
\end{table}

\begin{table}
  \begin{center}
    \begin{tabular}{l|cccc}
      & \multicolumn{4}{c}{Eigenvalues} \\ \hline

      $_\pm{\rm F}$ & 
      $(3\gamma-2)\sqz$ & 
      $-\frac{3}{2}(2-\gamma)\sqz$ &
      $-\frac{3}{2}(2-\gamma)\sqz$ &
      $\frac{3\gamma}{2}\sqz$ \\

      $_\pm{\rm K}$ &
      $4\sqz-2Q_+$ & 
      $3(2-\gamma)\sqz$ &
      0 &
      $3\sqz-\sqrt{\frac{3}{2}}\kappa U_0$ \\

      $_\pm\Phi$ &
      $-\frac{1}{2}(6-\kappa^2)\sqz$ &
      $-\frac{1}{2}(6-\kappa^2)\sqz$ &
      $-(2-\kappa^2)\sqz$ &
      $-(3\gamma-\kappa^2)\sqz$ \\

      $_\pm\Xi$ &
      $-3\frac{2+\kappa^2}{4+\kappa^2}\sqz$ &
      $-6\frac{\gamma+(\gamma-1)\kappa^2}{4+\kappa^2}\sqz$ &
      \multicolumn{2}{c}{
        $-\frac{3}{2}\frac{2+\kappa^2\pm\sqrt{(2+\kappa^2)(18-7\kappa^2)}}
        {4+\kappa^2}\sqz$} \\

      $_\pm{\rm FS}$ &
      $-\frac{3}{2}(2-\gamma)\sqz$ &
      $(3\gamma-2)\sqz$ &
      \multicolumn{2}{c}{
        $-\frac{3}{4\kappa}\left[(2-\gamma)\kappa
        \pm\sqrt{(2-\gamma)(24\gamma^2-(9\gamma-2)\kappa^2)}\right]\sqz$} \\

      $_\pm{\rm SSKS}$ & 
      $-3\frac{2-\gamma}{3\gamma-4}\sqz$ &
      $\frac{3\gamma}{3\gamma-4}\sqz$ &
      \multicolumn{2}{c}{
        $-\frac{3}{2}
        \frac{2-\gamma\pm\sqrt{(2-\gamma)(24\gamma^2-41\gamma+18)}}
        {3\gamma-4}\sqz$} \\ 
    \end{tabular}
    \caption{Eigenvalues for the different equilibrium points of the KS
      models.}\label{tab:eigenKS} 
  \end{center}
\end{table}

\begin{table}
  \begin{center}
    \begin{tabular}{lll}
      \multicolumn{3}{c}{Past attractors} \\ \hline
      Expanding from a singularity ($Q_0>0$)
      & $_+{\rm K}$ & $\kappa U_0<\sqrt{6}$ \\ \hline
      Contracting from a dispersed state ($Q_0<0$)
      & $_-\Phi$ & $\kappa^2<2$ \\ \hline\hline
      \multicolumn{3}{c}{Future attractors} \\ \hline
      Contracting to a singularity ($Q_0<0$)
      & $_-{\rm K}$ & $\kappa U_0<\sqrt{6}$ \\ \hline
      Expanding to a dispersed state ($Q_0>0$)
      & $_+\Phi$ & $\kappa^2<2$ \\ 
    \end{tabular}
    \caption{Summary of sources and sinks for the KS
      models.}\label{tab:stabKS} 
  \end{center}
\end{table}

The eigenvalues for each of the
equilibrium points are given in Table \ref{tab:eigenKS}.
The sources and sinks of the dynamical system when $\gamma>2/3$ are
listed in Table \ref{tab:stabKS} (all of the other equilibrium points
are saddles). Thus, there is always two segments
of the equilibrium set $_+{\rm K}$ that act as sources for
orbits. Similarly there are two segments on $_-{\rm K}$ that are
sinks. When $\kappa^2>2$, these are the only attractors, and all
solutions start from and recollape to a singularity ($K\rightarrow K$).

When $\kappa^2<2$, which then implies that $\kappa^2<3\gamma$ for $\gamma>2/3$,
the equilibrium points $_\pm\Phi$ are
attractors. Thus, for $\kappa^2<2$, there are also ever-expanding
($_+{\rm K}\rightarrow{_+\Phi}$) and ever-collapsing
($_-\Phi\rightarrow{_-{\rm K}}$) solutions.

From the expression for the monotonic function $M$ we deduce that {\em all} 
orbits asymptotically have $Q_+ \rightarrow 0$, $Q_0^2 \rightarrow 1$ or 
$\Omega_D \rightarrow 0$. Indeed, the existence of the monotonic function 
ensures that there are no periodic orbits and that generically orbits 
asymptote towards the local attractors (sinks and sources). Therefore, 
we can determine the global dynamics of the models.

To {\bf summarize}, when $\kappa^2>2$ all solutions start from and
recollape to a singularity ($K\rightarrow K$). Thus, in this case 
solutions can neither isotropize nor inflate. When $\kappa^2<2$, there are 
also ever-expanding (${\rm K}\rightarrow\Phi$) (and ever-collapsing
$\Phi\rightarrow{\rm K}$) solutions in addition to the recollapsing
solutions. Again, the $\Phi$ points correspond to power-law inflation
when $\kappa^2<2$. Consequently, in this case there is a subclass of
solutions that isotropize and inflate. 

The global asymptotic dynamics is similar to that in the case of
positive-curvature FRW models. However, due to the presence of shear,
the intermediate or transient dynamics can be quite different. In the
Kantowski-Sachs case the state space is  four-dimensional and so we
cannot display the phase portraits graphically  (as in the FRW
case). However, as an illustration we shall present the phase
portraits in the three-dimensional fluid vacuum and the massless
scalar field invariant sets in order to compare intermediate
behaviours.

\subsection{Fluid vacuum}\label{sec:vac}

\begin{figure}
 \centerline{\hbox{\epsfig{figure=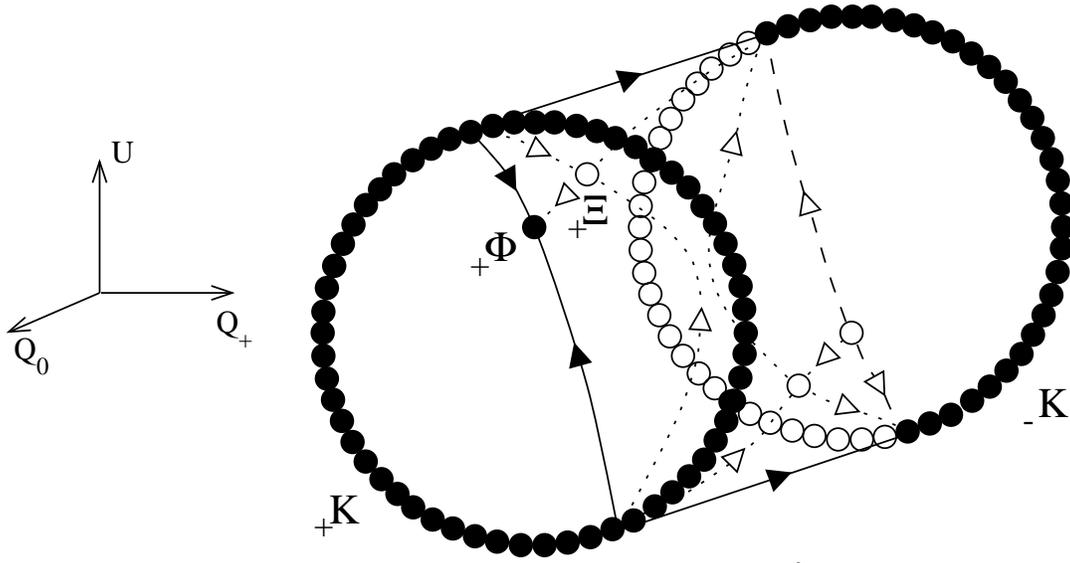,width=0.8\textwidth}}}
  \caption{The state space for Kantowski-Sachs models with no fluid 
    ($\Omega_D=0$), with $\kappa^2<2$. Dashed curves and white arrows and
  circles are screened. Dotted orbits lie in the $Z=0$ invariant
  submanifold.}\label{fig:KSvac1}  
\end{figure}

\begin{figure}
  \centerline{\hbox{\epsfig{figure=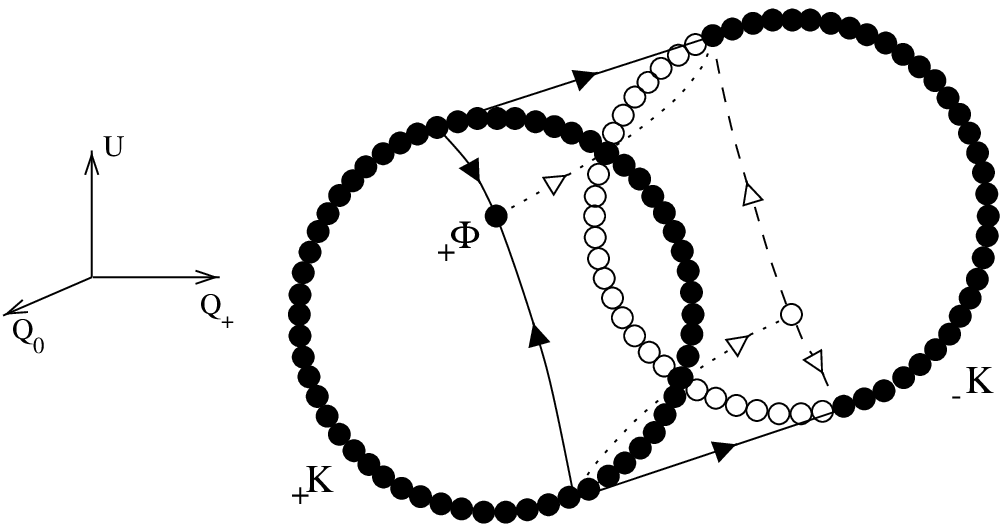,width=0.8\textwidth}}}
  \caption{The state space for Kantowski-Sachs models with no fluid
    ($\Omega_D=0$), with $2<\kappa^2<6$. Dashed curves and white arrows and
  circles are screened. Dotted orbits lie in the $Z=0$ invariant
  submanifold.}\label{fig:KSvac2} 
\end{figure}

\begin{figure}
  \centerline{\hbox{\epsfig{figure=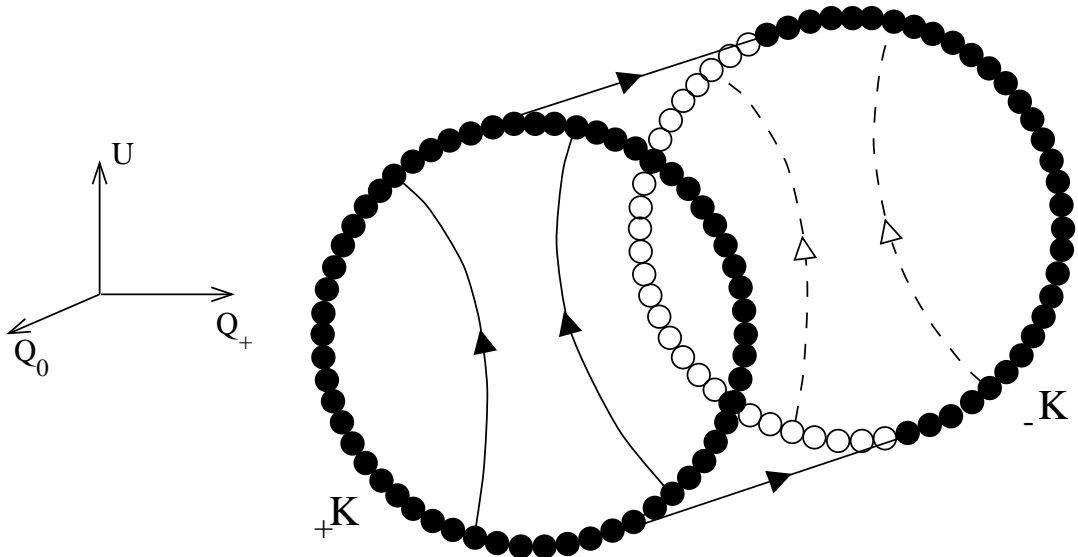,width=0.8\textwidth}}}
  \caption{The state space for Kantowski-Sachs models with no fluid
    ($\Omega_D=0$), with $\kappa^2>6$. Dashed curves and white arrows and
  circles are screened.}\label{fig:KSvac3} 
\end{figure}

The fluid vacuum ($\Omega_D=0$) is an invariant submanifold, as seen from 
Eq. (\ref{eq:Om}). Using the Friedmann equation to eliminate $W$, 
we obtain a three-dimensional dynamical system in $(Q_0,Q_+,U)$:
\begin{eqnarray}
  Q_0'&=&(1-Q_0^2)\left[1+Q_0Q_+ -3(Q_+^2+U^2)\right] , \\
  Q_+'&=&-(1-Q_0^2)(1-Q_+^2)-3Q_0Q_+(1-Q_+^2-U^2) , \\
  U'&=&(1-Q_0^2)Q_+U+
  \left(\sqrt{\frac{3}{2}}\kappa-3Q_0U\right)(1-Q_+^2-U^2) .
\end{eqnarray}
From table \ref{tab:qKS}, it is immediately seen that the equilibrium 
points that are contained in this submanifold are $_\pm{\rm K}$,
$_\pm\Phi$, and $_\pm\Xi$. The state space is depicted in Figs.
\ref{fig:KSvac1}, \ref{fig:KSvac2} and \ref{fig:KSvac3}.
Note that $Z=0$, where $Z$ is defined by
\begin{equation}
  Z=Q_+-\frac{1}{2}Q_0+\frac{\sqrt{6}}{2\kappa}U ,
\end{equation}
is an invariant submanifold, and that both $_\pm\Phi$ and $_\pm\Xi$ are
contained in this submanifold. 

\subsection{Massless case}\label{sec:massless}

\begin{figure}
  \centerline{\hbox{\epsfig{figure=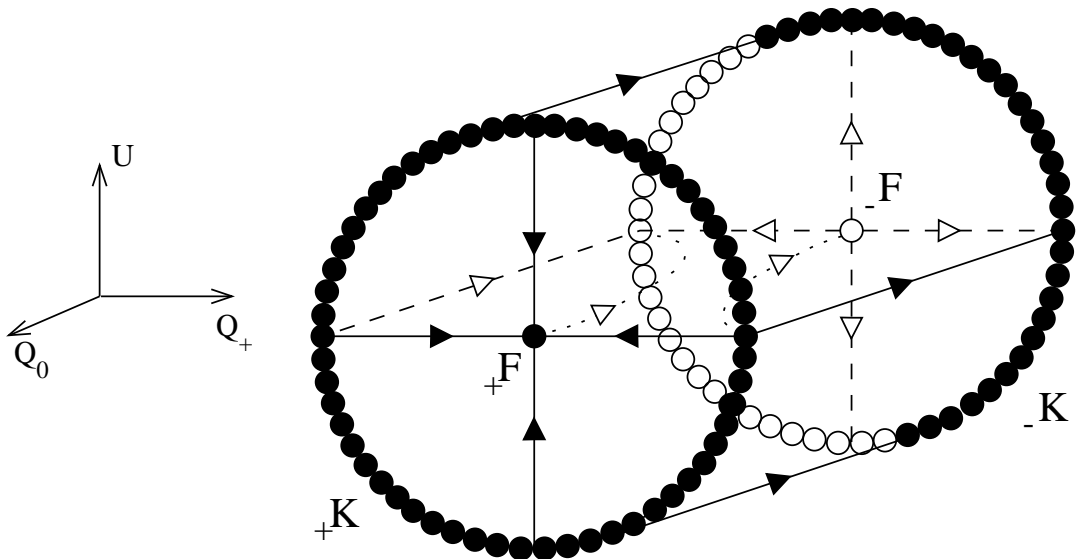,width=0.8\textwidth}}}
  \caption{The state space for Kantowski-Sachs models with fluid and 
    massless scalar field, $W=0$. Dashed curves and white arrows and
    circles are screened. The horizontal plane is the $U=0$ invariant
    submanifold (compare with \protect\cite{SSS}).}\label{fig:KSmassless} 
\end{figure}

The massless case corresponds to the invariant submanifold
$W=0$, which leads to a three-dimensional system in ($Q_0,Q_+,U$):
\begin{eqnarray}
  Q_0'&=&-(1-Q_0^2)\left[\frac{3\gamma-2}{2}-Q_0Q_+ 
  +\frac{3}{2}(2-\gamma)(Q_+^2+U^2)\right] , \nonumber \\
  Q_+'&=&-(1-Q_0^2)(1-Q_+^2) -\frac{3}{2}(2-\gamma)Q_0Q_+\Omega_D ,  \\
  U'&=&   U\left[(1-Q_0^2)Q_+-\frac{3}{2}(2-\gamma)Q_0\Omega_D\right] .
  \nonumber 
\end{eqnarray}
From table \ref{tab:equiKS}, it is immediately seen that the equilibrium 
points that are contained in this submanifold are $_\pm{\rm K}$
and $_\pm{\rm F}$. The state space is depicted in Fig. \ref{fig:KSmassless}.

\section{Discussion}

We have studied closed cosmological models with a perfect fluid
satisfying a linear equation of state with $2/3<\gamma<2$ and a scalar
field with an exponential potential. We have utilized a new set of
normalised variables which lead to the compactification of state
space, enabling us to apply the theory of dynamical systems to
determine the qualitative properties of the models. In all cases we
have been able to find monotonic functions which, together with a
local analysis of the equilibrium points, enable us to determine the
global properties of the models. 

We first studied the closed FRW cosmological models. We found that
when $\kappa^2>2$, all solutions start from and recollape to a
singularity ($K\rightarrow K$). In this case solutions generically do
not inflate. When $\kappa^2<2$, solutions can either
recollapse ($K\rightarrow K$) or expand forever
(${\rm K}\rightarrow\Phi$) towards power-law inflation solutions (or
collapse forever $\Phi\rightarrow{\rm K}$); consequently, in this case
there is a subclass of solutions that inflate. A number of phase
portraits were displayed. 
 
These results generalise previous qualitative work on
positive-curvature FRW models with a scalar field (only) \cite{R3} and
with a scalar field plus a barotropic perfect fluid \cite{HCW} in
which compactified variables were not utilized, and rigorous analyses
of perfect fluid (only) models using compactified variables
\cite{WE,Aber}, and completes and generalises more recent work using
different compactified variables \cite{turner}. We also note that
positive-curvature FRW models with a perfect fluid and a positive
cosmological constant have been investigated recently using
qualitative methods and utilizing compactified variables \cite{G}.

In the case of the Kantowski-Sachs models we again found that when 
$\kappa^2>2$ all solutions start from and recollape to a singularity
($K\rightarrow K$) and can consequently neither isotropize nor
inflate. When $\kappa^2<2$, there are also ever-expanding
(${\rm K}\rightarrow\Phi$) (and ever-collapsing
$\Phi\rightarrow{\rm K}$) solutions in addition to the recollapsing
solutions, where again the $\Phi$ points correspond to the flat FRW
power-law inflationary solution. Consequently, in this case there is a
subclass of solutions that isotropize and inflate. 

The investigation of Kantowski-Sachs models complements the study of
Bianchi models \cite{bchio} and completes the analysis of spatially
homogeneous models. Collins \cite{collins} studied perfect fluid
Kantowski-Sachs models qualitatively using expansion-normalised
variables (for which the state space was non-compact) and showed that
all models start at a Big Bang and recollapse to a final ``Big Crunch''
singularity. This work was generalised recently by Goliath and Ellis
\cite{G} in which Kantowski-Sachs models with a perfect fluid and a
cosmological constant were investigated using qualitative methods and
utilizing the compactified variables of Uggla and Zur-Muhlen
\cite{Uggla-zurMuhlen}; particular attention was focussed upon whether
the models isotropize, thereby explaining the presently observed
near-isotropy of the universe. More importantly, Kantowski-Sachs
models with a scalar field and an exponential potential, but without
barotropic matter, have been studied qualitatively \cite {R5},
although compactified variables were not utilized. 

To conclude an analysis of positive-curvature spatially homogeneous
cosmological models with a perfect fluid and a scalar field with an
exponential potential, Bianchi type IX models would need to be
studied. However, such a study is beyond the scope of the current
paper. For example, Bianchi type IX models are known to have
very complicated dynamics, exhibiting the characteristics of chaos
\cite{WE,chaos}. However, partial results are known. Bianchi type IX models
with a scalar field (only) have been studied qualitatively, with an
emphasis on whether these models can isotropize \cite{HO}. Scalar-field
models with matter have also been studied \cite{R9}. For
example, it has been shown that the power-law inflationary solution is
an attractor for all initially expanding Bianchi type IX models except
for a subclass of the models which recollapse
\cite{R9,ColeyIbanezVanDenHoogen}. However, compact variables have not
been utilized and the analyses were not rigorous. 

A more rigorous treatment of the class of Bianchi type IX models with
a non-tilted perfect fluid (only) using compactified variables has
been possible \cite{WE}. Although an appropriately defined normalised
Hubble variable is found to be monotonic, enabling some results to be
obtained, several problems remain open. More rigorous global results
are possible. For example, Bianchi type IX models with matter have
been shown to obey the ``closed universe recollapse'' conjecture
\cite{LW}, whereby initially expanding models enter a contracting
phase and recollapse to a future ``Big Crunch''. In addition, Ringstr\"om
has proven that a curvature invariant is unbounded in the
incomplete directions of inextendible null geodesics for generic
vacuum Bianchi models \cite{ring}, and rigorously shown that the
Mixmaster attractor is the past attractor of Bianchi type IX models
with an orthogonal perfect fluid \cite{ring2}. A complete qualitative
analysis of the special class of locally rotationally symmetric
Bianchi type IX perfect fluid models, which do not exhibit oscillatory
or chaotic behaviour near to the initial or final singularities, has
been given in \cite{Uggla-zurMuhlen}, based upon an appropriately
defined set of bounded variables. 

The Kantowski-Sachs models exhibit similar global properties to the
positive-curvature FRW models; in particular, for $\kappa^2>2$ all
initially expanding models reach a maximum expansion and thereafter
recollapse, whereas for $\kappa^2<2$ models generically recollapse or
expand forever towards a flat isotropic power-law inflationary
solution. The Bianchi type IX models share these qualitative
properties. However, the intermediate behaviour of the Kantowski-Sachs
models can be quite different to that of the FRW models. In order to
illustrate the possible intermediate dynamics of the Kantowski-Sachs
models, we studied the special cases of no barotropic fluid, and a
massless scalar field in Secs. \ref{sec:vac} and \ref{sec:massless},
respectively (see Figs. \ref{fig:KSvac1} -- \ref{fig:KSmassless}). 

Finally, we remark that the dynamics of the Kantowski-Sachs models
obtained here will be crucial for understanding the global dynamics of
general self-similar spherically symmetric models \cite{CG}.

\section{Acknowledgements}

We would like to thank Peter Turner for his help with the analysis of
the positive-curvature FRW models \cite{turner}. AC would like to
acknowledge financial support from NSERC of Canada.
MG would like to thank the Department of Mathematics and Statistics
at Dalhousie University for hospitality while this work was carried out.

\end{document}